# A Workflow for Exploring Ligand Dissociation from a Macromolecule: Efficient Random Acceleration Molecular Dynamics Simulation and Interaction Fingerprints Analysis of Ligand Trajectories.

Daria B. Kokh[a*], Bernd Doser [b], Stefan Richter[a], Fabian Ormersbach[a], Xingyi Cheng [a, c], Rebecca C. Wade[a,d,e*]

[a]Molecular and Cellular Modeling Group, Heidelberg Institute for Theoretical Studies, Schloss-Wolfsbrunnenweg 35, 69118 Heidelberg, Germany

[b]Heidelberg Institute for Theoretical Studies, Schloss-Wolfsbrunnenweg 35, 69118 Heidelberg, Germany

[c]Molecular Biosciences, Heidelberg University, Im Neuenheimer Feld 282, 69120, Heidelberg, Germany

[d]Center for Molecular Biology (ZMBH), DKFZ-ZMBH Alliance, Heidelberg University, Im Neuenheimer Feld 282, 69120 Heidelberg, Germany

[e]Interdisciplinary Center for Scientific Computing (IWR), Heidelberg University, Im Neuenheimer Feld 205, Heidelberg, Germany

[*]Daria.Kokh@h-its.org, Rebecca.Wade@h-its.org

**ABSTRACT**

The dissociation of ligands from proteins and other biomacromolecules occurs over a wide range of timescales. For most pharmaceutically relevant inhibitors, these timescales are far beyond those that are accessible by conventional molecular dynamics (MD) simulation. Consequently, to explore ligand egress mechanisms and compute dissociation rates, it is necessary to enhance the sampling of ligand unbinding. Random Acceleration MD (RAMD) is a simple method to enhance ligand egress from a macromolecular binding site, which enables the exploration of ligand egress routes without prior knowledge of the reaction coordinates. Furthermore, the τRAMD procedure can be used to compute the relative residence times of ligands. When combined with a machine-learning analysis of protein-ligand interaction fingerprints (IFPs), molecular features that affect ligand unbinding kinetics can be identified. Here, we describe the implementation of RAMD in GROMACS 2020, which provides significantly improved computational performance, with scaling to large molecular systems. For the automated analysis of RAMD results, we developed MD-IFP, a set of tools for the generation of IFPs along unbinding trajectories and for their use in the exploration of ligand dynamics. We demonstrate that the analysis of ligand dissociation trajectories by mapping them onto the IFP space enables the characterization of ligand dissociation routes and metastable states. The combined implementation of RAMD and MD-IFP provides a computationally efficient and freely available workflow that can be applied to hundreds of compounds in a reasonable computational time and will facilitate the use of τRAMD in drug design.

## I. INTRODUCTION

Many crystal structures of bound ligand-protein complexes reveal that small molecules are often positioned in a cavity that is completely or partially buried in the protein, where no clear entrance or exit route can be observed. This suggests that macromolecular conformational rearrangements associated with opening and closing of entrance/exit channels or tunnels are required for ligand binding and unbinding. While such rearrangements are dependent on the mobility of the macromolecule itself, the ligand may influence this motion, which results, in particular, in variations in the binding kinetics as well as in the binding pocket shape when different ligands bind. Characterization of the dissociation of protein-ligand complexes can give insights into binding and unbinding mechanisms and support the design of new therapeutic agents aimed at modulating protein function[1,2]. Besides, kinetic parameters, such as the unbinding rate, can have a critical impact on the *in vivo* drug efficacy. The physical process of ligand unbinding has proven difficult to elucidate as it usually occurs on a timescale that is beyond the simulation times feasible in conventional molecular dynamics (MD) simulations. Amongst the different strategies for reducing the computational time required to observe ligand unbinding events (such as metadynamics [3,4,5,6] or weighted ensemble[7,8] simulations), a number of nonequilibrium MD simulation methods have been found to be computationally efficient and to enable the estimation of the kinetics of unbinding processes and the elucidation of the features governing unbinding[9–11,12,13,14,15].

One approach is to facilitate ligand unbinding from a binding site in a macromolecule by applying an additional small, randomly oriented force to the center of mass of the ligand during otherwise conventional MD simulations of a solvated protein-ligand complex. This method, originally called Random Expulsion Molecular Dynamics[16] and later referred to as Random Acceleration MD (RAMD) because of its more general scope of application, was designed to identify possible ligand egress routes from a buried binding site . In this approach, the magnitude of the additional ligand force or acceleration is kept constant during the simulation, while the orientation of the force or acceleration is chosen randomly. The displacement of the ligand center of mass relative to the starting position is checked after defined time intervals (usually 100 fs) and either a new force direction is chosen randomly if the ligand displacement is below a threshold distance, or the force direction is kept for the next time interval if the ligand displacement is above the threshold distance. This process is repeated until the ligand displacement exceeds a predefined distance from its starting position, at which point it is considered to have dissociated from the macromolecule.

RAMD was first implemented in the ARGOS program[16,17,18] and applied to explore the dissociation routes of a set of substrates of cytochrome P450 enzymes from their buried active sites.[16,19,20,21] The validity of the ligand routes from cytochrome P450cam explored by RAMD in Ref. [16] has been shown later in unbiased MD simulations (57 μs in total) of substrate binding[22]. The method has since been used in multiple studies, including its implementation in AMBER8[21] for studying cytochrome P450 enzymes[23] and G-protein coupled receptors (GPCR) (rhodopsin[24], β2-adrenergic receptor[25]) , in GROMACS[26] for studying ligand dissociation from liver fatty acid binding protein[27], and in CHARMM[28] for unbinding simulations of retinoic acid from the retinoic acid receptor[21]. The need to implement RAMD in a way that was sustainable in constantly upgraded MD software became apparent. This problem was solved by the implementation of RAMD as a tcl script wrapping the NAMD[29] package in Ref.[30,31] The NAMD implementation of RAMD hardly needed any additional adaptation even though the NAMD package was constantly developed and updated. However, this implementation has shown limited performance due to the intrinsic bottleneck in its parallelization, as after each MD time interval to assess ligand motion and decide on changing the force direction, the simulation has to be paused. This limiting step hinders the application of the method to larger systems, such as membrane proteins or protein complexes, whose simulation speed has particularly benefited from recent improvements in the efficiency of MD simulation code parallelization.

Recently, the NAMD implementation of RAMD was further revised by reverting to using a force instead of an acceleration as the input parameter with the aim of estimating relative ligand residence times[32]. Employing a constant force magnitude in simulations for a series of different ligands ensures independence of the perturbation effect on the ligand mass. In this case, the residence times obtained for different compounds can be compared. For this purpose, the time required for the ligand to leave the binding pocket is computed in multiple RAMD dissociation trajectories starting from several different snapshots, i.e. coordinates and velocities, from MD equilibration runs. The dissociation times are then used to derive relative residence times. This protocol, τRAMD, was evaluated on more than 90 inhibitors of heat shock protein 90 (HSP90), and showed a good correlation between computed relative residence times and experimental data[32,33]. In Ref. [33], it was also demonstrated that the simulated dissociation trajectory can be further analyzed to decipher the molecular determinants that affect ligand residence time by computing and statistically analyzing protein-ligand interaction fingerprints, IFPs, for the parts of the dissociation trajectories in which the ligands are egressing from the binding site. Each IFP represents a 3D protein-ligand interaction profile by a binary vector of interactions (such as hydrogen- or halogen-bonds, aromatic stacking, salt bridges, or hydrophobic contacts) defined by the distance- and angle- thresholds specified for each type of the interaction. In Ref.[33], OpenEye's OEChem Toolkit[34] was employed to compute IFPs.

In the present paper, we first report a new implementation of the RAMD method in the GROMACS 2020 MD simulation package[35] as a part of the PULL function. This implementation enables simulation time to be decreased by more than ten times, depending on the system size, compared to the NAMD implementation. We tested the performance of the new implementation on two target proteins: the N-terminal domain of HSP90, a rather small soluble protein that has been the subject of a number of studies of ligand dissociation.[36–39], and the muscarinic receptor M2, a GPCR embedded in a lipid bilayer (explored using metadynamics simulations in Refs. [4,40]).

Secondly, we present an open-source software tool set, MD-IFP, for the computation of protein-ligand IFPs along simulated MD trajectories. We evaluate the use of MD-IFP on a set of over 40 protein-ligand complexes and then demonstrate its application to analysis of RAMD simulations of the complexes of HSP90 with three inhibitors that have different residence times and distinct binding mechanisms. Although there are several freely available software tools for the computation of the three-dimensional structural protein–ligand interaction fingerprints from the coordinates of protein-ligand complexes (in PDB format), such as SPLIF[41], PLIP[42], FLIP[43], and LIGPLOT[44], none of them is, to the best of our knowledge, designed to be integrated as a part of automated protocol for MD trajectory analysis. Furthermore, we developed an approach to the analysis of the computed IFPs, which enable the dissociation routes to be classified and metastable states for ligand dissociation to be detected. To evaluate the workflow presented in this manuscript, we compared simulation results of the new protocol with the previous τRAMD implementation, and evaluated the accuracy of MD-IFP computations and the ability of the analysis to reveal details of dissociation mechanisms. We show that the workflow facilitates effective analysis of the main ligand dissociation pathways from the protein binding site and identification of the molecular characteristics of metastable states along these pathways.

**THEORY**

**A. The RAMD procedure and the choice of MD simulation parameters.**

In the RAMD approach, a small randomly oriented force of the constant magnitude is applied to the center of mass, COM, of the ligand during MD simulations. The orientation of the force is chosen randomly at the beginning and, after a short simulation interval, its orientation is changed randomly if the displacement of the ligand COM relative to the protein COM is below a specified threshold or retained otherwise. Simulation intervals are repeated until the ligand COM relative to the protein COM reaches a pre-defined distance, when the ligand is considered to have dissociated from its binding partner. There are several parameters that have to be assigned in this procedure: (i) the time interval for checking the motion of the ligand COM and deciding whether to change the direction of the random force, (ii) the COM displacement threshold defining whether the direction of the force should be changed, (iii) the maximum displacement of the COM indicating when simulations are stopped, and (iv) the magnitude of the force applied. Additionally, the standard MD simulation parameters, such as the type of thermostat (and barostat) must be selected. The relaxation time parameter of the thermostat can strongly affect the ligand dissociation time, since it influences the dissipation of the additional kinetic energy of the ligand due to the external random force.

The maximum displacement of the ligand COM is defined by the approximate protein extent from the protein COM plus 5-10 Å. A larger distance is better than a smaller one since, as soon as the ligand interaction with the protein is lost, its motion is driven by the external force and becomes very fast and any reasonable threshold can be reached within a few simulation steps. Altering any of the other parameters leads to a change in dissociation time, but their effects can compensate each other. For example, a larger force magnitude leads to faster dissociation, whereas a longer COM displacement threshold makes dissociation slower.   Thus, there is no single choice of suitable MD parameters, but in order to be able to evaluate the relative ligand residence times in the τRAMD procedure[32], the parameters must be kept constant in all the simulations of a set of ligands that are to be compared. The most obvious criterion for the parameter fitting would be the best agreement of the simulation results with experimental residence times for a set of compounds. However, given the limited data on experimental residence times, it is generally difficult to make an unambiguous choice. Therefore, in the τRAMD procedure, the values of all parameters except one, the random force magnitude, are fixed for all systems studied.  Specifically, for simulations using the NAMD software, the Langevin thermostat and barostat are used, both with a relaxation time of 1ps. In the GROMACS implementation, the Nosé–Hoover thermostat and Parrinello-Rahman barostat are employed (an analysis of the GROMACS simulation parameters is given the Section Results A. Benchmark of the GROMACS 2020 implementation of RAMD). The length of the time interval for assessing ligand motion and force direction is set to 100 fs and the COM displacement threshold is set to 0.025 Å.  While these parameters can be used for all protein-ligand systems studied, the magnitude of the random force may need to be adjusted according to the properties of the protein-ligand complexes studied. It is recommended to use a value of 14 kcal/molÅ for the initial simulations and to adjust this value if too few egress events or too fast dissociations are observed. The criteria for the choice of the random force magnitude and its adjustment are given in Ref.[32]

### B. τRAMD protocol

The τRAMD  protocol was reported in Refs. [32,33] . Here, we briefly outline the main steps. A set of starting snapshots (at least four replicas) is generated using conventional MD simulations, ideally from several independent trajectories (started with different coordinates and/or velocities). Each starting snapshot is then used to generate a set of 15-30 RAMD ligand dissociation trajectories.  The effective residence time for each starting replica is defined by the dissociation time, corresponding to 50% of the cumulative distribution function (CDF) for the set of RAMD trajectories as illustrated in **Fig. 1 A**. A bootstrapping procedure (5000 rounds with 80% of samples selected randomly) is performed to obtain a residence time for each replica, $\tau_{repl}$, which should converge to a Gaussian-like distribution if the sampling is sufficiently large (**Fig. 1 B**). Otherwise, the number of simulated trajectories for this replica should be increased. Additionally, a Kolmogorov-Smirnov (KS) test can be done to assess the sampling quality (**Fig. 1 C**). The final relative residence time, $\tau_{RAMD}$, is defined as the mean of $\tau_{repl}$ over all replicas (**Fig. 1 D**).

### C. Protein-ligand interaction fingerprints to characterize simulated ligand egress trajectories

We considered the following classes of receptor-ligand interactions to define the IFPs in the MD-IFP analysis: hydrophobic (HY), aromatic (AR), hydrogen bond donor (HD) or acceptor (HA), salt bridge (IP/IN), halogen bonds (HL), and water bridge (WB). Parameters of IFPs computed in each class are summarized in **Table 1**. The identification of the receptor-ligand IFPs is done using:  (i) the RDKit[45] software, which identifies the chemical properties of the ligand using an input ligand mol2 format file, and (ii) the MDAnalysis[46] python library. The latter library enables the reading of MD trajectories, independently of their format or the program used to generate them, as well as frame-by-frame operations on the trajectories (such as computing distances and detecting hydrogen bonds between selected groups of atoms). Prerequisites for the IFP calculation are a structure of a protonated protein-ligand complex and a ligand structure in mol2 format. For the detection of water bridges, energy minimization of the structure of the protein-ligand complex is desirable.

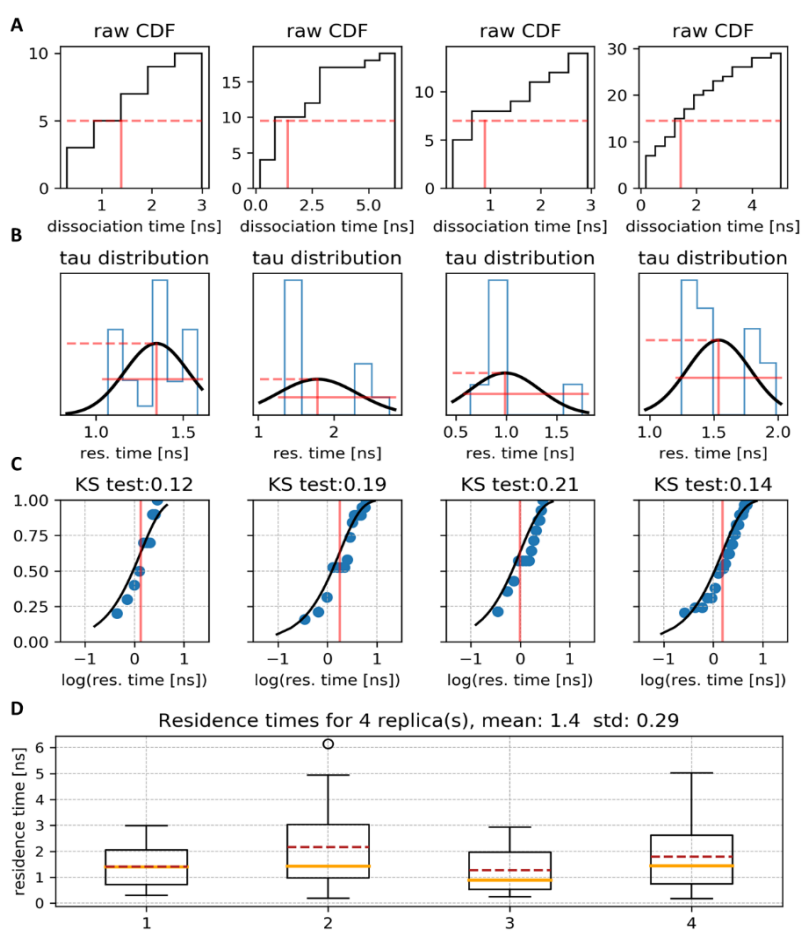

.

*Figure 1 Use of the τRAMD method to compute the residence time of an inhibitor of HSP90, compound 16, PDB ID: 5J86 (the NAMD implementation was used, the same plots for the Gromacs implementation under different thermostat conditions for all ligands are shown in Figs. S1)* : **(A)** *Cumulative distribution function (CDF) for four sets of 15 RAMD dissociation times, with each trajectory in a set starting from the same replica (i.e. the same starting coordinates and velocities); the effective residence time obtained from these raw data is indicated by the red solid line at the time when the ligand has dissociated in half of the trajectories (1/2 of the maximum value of the CDF, which is indicated by the red dashed line);* **(B)** *Distribution function of effective residence times obtained after bootstrapping of the raw data along with the corresponding Gaussian distribution (black line); the mean (log($\tau_{repl}$)) and half-width are indicated by red lines;* **(C)** *Poisson cumulative distribution function (PCDF, black line), $P = 1 - \exp(-t/\tau_{repl})$ is compared with the empirical cumulative density function (ECDF, blue points) obtained from the dissociation probability distribution; $\tau_{repl}$ is indicated by the red line; the results of the Kolmogorov-Smirnov (KS) test are quantified by the supremum of the distance D between the Poisson and empirical CDFs (denoted above the plot);* **(D)**

*Bar plot of the relative residence times obtained from each replica. The box extends from the lower to upper quartile values of the data, the whiskers show the range of the data, outliers are shown by points, the median and mean are shown by orange and dashed red lines, respectively. The average residence time in ns computed from all replicas is shown above the plot.*

*Table 1 Definitions of interactions used to compute IFPs in MD-IFP analysis.*

|   | Type of interaction | Ligand atom type or atom name | Protein atoms | Criteria: maximum distance [a] (in Å) or angle (in degrees) |
|---|---|---|---|---|
| HY | Hydrophobic[b] | Hydrophobic or Fluorine | S of MET or CYS<br>Carbons except for: CG in ASN ASP; CD in GLU GLN ARG; CZ in TYR ARG; CB in SER THR; CE in LYS; C in backbone | 4.0[47] |
| IP | Salt bridges | Pos.Ionizable | OE* OD* of ASP GLU | 4.5[48] |
| IN | Salt bridges | Neg.Ionizable | NH* NZ in ARG LYS<br>HD HE in HI2 | |
| AR | $\pi$ - stacking [c] | Aromatic | atoms CZ* CD* CE* CG* CH* NE* ND* in PHE TRP TYR HIS HI2 HIE HID[c] | 5.5 [47] |
|    | Cation - $\pi$ [d] | PosIonizable | atoms CZ* CD* CE* CG* CH* NE* ND*[c] in PHE TRP TYR HIS HI2 HIE HID | 5.0 [49] |
|    | $\pi$ - Cation [d]<br>$\pi$ - Amide | Aromatic | Nitrogens in ARG an LYS;<br>backbone nitrogen; | 5.0[49]<br>5.0[50] |
| HL | Halogen - $\pi$ | Chlorine, Bromine, Iodine [c] | atoms CZ* CD* CE* CG* CH* NE* ND* ins HE TRP TYR HIS HIE HID | 3.5[51] |
|    | Halogen-carboxyl | | O in the backbone or in ASP GLU; | 3.5[51] |
|    | Halogen-S | | S in MET and CYS [c] | 3.5[52] |
| HD | Hydrogen bond | Donor | O OC1 OC2 OH2 OW OD1 OD2 SG OE1 OE1 OE2 ND1 NE2 SD OG OG1 OH[f] | 3.3, 100° |
| HA | Hydrogen bond | Acceptor[e] | N OH2 OW NE NH1 NH2 ND2 SG NE2 ND1 NZ OG OG1 NE1 OH[f] | |
| WB | Water bridge between protein and ligand | Donor/Accept | default donor/acceptor as for H-bonds | 3.3, 100° |

[a] Distance between heavy atoms for all interactions except for the H-bond, where the distance between the hydrogen and acceptor atoms is considered

[b] Hydrophobic atoms in proteins : all carbon atoms of the protein except those bound with a double bond to oxygen (C=O) or nitrogen (C=N) (i.e. backbone C atom, carboxy C in ASP and GLU, amide C in GLN and ASN, guanidino C in ARG), and with a single bond to OH or NH3 (as in SER, TYR, LYS and THR side chains)

[c] Contact with at least 5 atoms from the list (angle is not considered)

[d] The cation-pi interaction is orientation-dependent with the strongest interaction when the cation is placed next to an aromatic ring; to take into account the case of trimethylamine (ligand), we use a slightly longer distance threshold in the case of a ligand cation.

[e] As defined in RDkit; fluorine is also considered as an acceptor. In RDKit, the definitions of the feature types (Donor, Acceptor, Aromatic, Halogen, Basic, and Acidic) were adapted from[53]

[f] as defined in the Charmm force field used in MDAnalysis

Hydrophobic contacts (HY) were defined between the ligand hydrophobic carbon atoms as identified by RDKit (carbon atoms adjacent to any O or N are not considered as hydrophobic) or the ligand fluorine and the protein carbon (or sulfur) atoms at an inter-heavy atom distance of up to 4.0 Å (with an implicit treatment of hydrogen atoms). The definition employed is consistent with the definition used by the PLIP[42] server, but different from, for example, OpenEye's OEChem Toolkit[34], where all carbon atoms are considered to be hydrophobic.

Hydrogen bonds (HD/HA) were defined by an acceptor-hydrogen distance $\leq$ 3.3 Å[47], and a donor-hydrogen-acceptor angle $\geq$ 100°. A smaller angle is usually chosen to make the definition less strict, which is sometimes necessary for the detection of H-bonds in less accurate crystal structures or in MD trajectories.

The aromatic interaction class (AR) includes $\pi$-$\pi$ interactions (both plane and edge[54,47]), cation -$\pi$, and amide-$\pi$ interactions. Importantly, all aromatic interactions were defined solely on the basis of interatomic distances. The mutual orientation of the interacting fragments was taken into account implicitly by setting the minimum number of non-hydrogen contact atoms to be at least 5 (illustrated in **Fig. S2**). This definition is, therefore, less strict than, for example, in PLIP[42] where, for the detection of $\pi$-$\pi$ stacking, the angle between the two $\pi$ rings was defined by a threshold. Nevertheless, with this definition, only a few cases were detected where the aromatic rings are slightly tilted relative to the in-plane and edge position (see method evaluation below).

Halogen bonds include interactions between a polarized halogen (not fluorine) atom (Cl, Br, I, as a donor), sulfur, and a nucleophile or an aromatic ring (acceptor). The interaction distance and orientation of this type of bond varies across interaction partners with a rather wide distribution of the possible values [51]. For halogen-aromatic interactions we employed an average distance threshold of 3.5 Å without taking into account the mutual orientation of the interacting fragments but using a threshold of 5 contact atoms. For halogen-carboxyl interactions the C-Hal…O angle should be above 170°. We therefore count only halogen-carboxyl interactions if there are no ligand carbon atoms within the distance of Hal…O plus 1 Å from the oxygen atom.

Salt bridges were split into two classes: those with positive (IP) and those with negatively (IN) ionizable ligand atoms with a maximum distance from the respective protein heavy atoms of 5.0 Å. This assignment was based on a recent analysis of protein-ligand contacts in crystal structures[48] (the acceptor-hydrogen distance in salt bridges was found to be within 2.8-3.3 Å). Note that this distance threshold is smaller than the threshold of 5.5 Å suggested in Ref. [55] and employed, for example, in PLIP[42].

Protein-ligand water bridges were identified based on the h-bond detection function of the MDAnalysis package (the same H-bond parameters, 3.3Å and 100° were employed).

Finally, all nonspecific protein-ligand contacts within a threshold of 5Å between heavy atoms were stored, along with the number of water molecules in the ligand solvation shell defined by a threshold of 3.5 Å between ligand and water heavy atoms.

**D. Trajectory analysis**

IFPs were computed for each of the last 300 snapshots of each RAMD dissociation trajectory with the snapshots being saved with a stride of 2 ps. This part of the trajectories comprises a short sampling of the bound state and the ligand dissociation phase for the major of trajectories (see discussion below). All structures were superimposed with a single reference structures obtained after initial equilibration of the system. Additionally, to IFPs, the number of water molecules in the first water shell around the ligand, the root mean squared deviation (RMSD) from the bound position of the ligand, and the ligand centre of mass, COM, were computed for each snapshot.

For each MD snapshot, a binary IFP vector was stored that contains either 0 or 1 for each contact showing the presence or absence of a particular interaction. A complete set of IFPs for all snapshots, each represented as a binary vector, was combined into a single matrix (with frames along the trajectory as rows and IFP vector elements as columns) for the further analysis.

We employed kmeans clustering to identify the states most often visited in the IFP (see details of the clustering procedure the Appendix E, Analysis Protocol). The positions of the ligand in all frames that belong to a particular cluster can be projected onto physical space by mapping the ligand COMs onto a 3D grid and summing over all snapshots in the cluster. Importantly, the COM distribution in a cluster may not be compact and different ligand orientations with close COMs may be assigned to different clusters. The dissociated state is defined by the cluster in which no protein-ligand contact is found or in which multiple non-specific contacts are present, with the ligand COM spread around the protein. In contrast, the clusters describing the bound states of the ligands are usually compact in the physical space.

## II. RESULTS AND DISCUSSION

### A. Benchmark of the GROMACS 2020 implementation of RAMD

We benchmarked the GROMACS 2020 implementation of RAMD and its use in the τRAMD protocol for three inhibitors of HSP90 (compounds 8, 16 and 20 with $k_{off}$ = 0.21 $s^{-1}$, 1.4 $10^{-2}$ $s^{-1}$ and 1.4 $10^{-4}$ $s^{-1}$, respectively, studied in Ref. [36] ; PDB ID: 5J64, 5K86, and 5LQ9,) selected to have distinct binding scaffolds (**Fig.2**) and large differences in their residence time.

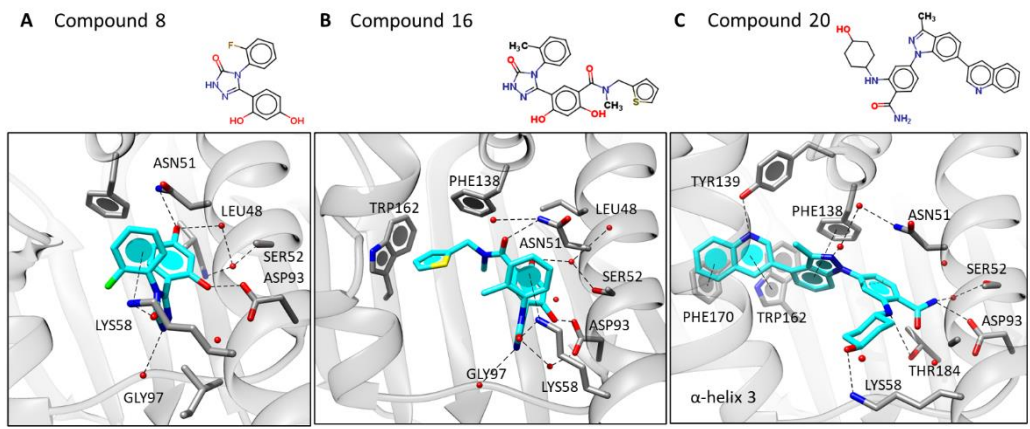

*Figure 2*: Illustration of the binding site of HSP90 with three inhibitors bound (A, B, C - compounds 8, 16 , and 20 of Ref. [36], PDB ID: 5J64, 5J86, 5LQ9, respectively). These protein-ligand complexes were employed for the evaluation of the GROMACS 2020 implementation of RAMD. The ligands are shown with cyan carbons and the protein is shown in half-transparent cartoon representation with interacting residues in stick representation; water molecules are indicated by red spheres; hydrogen bonds and aromatic interactions are denoted by dashed lines.

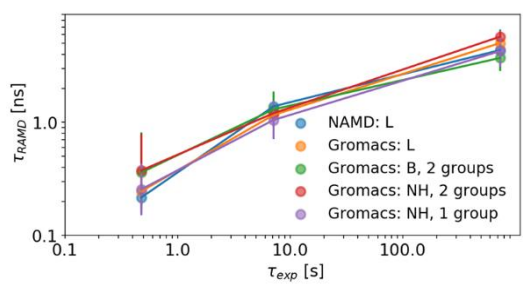

*Figure 3* Relative residence time, $\tau_{RAMD}$, computed for three inhibitors (compounds 8, 16 and 20) of HSP90 and for different simulations conditions plotted against measured residence time. The NAMD implementation from Ref.[32] and the GROMACS 2020 implementation described here were used. In NAMD simulations Langevin thermostat was used, in simulations using the Gromacs engin– Berendsen (B) and Nosé–Hoover (NH) thermostats as well as Langevn (L) dynamics were tested. Additionally, Gromacs simulations with thermostat applied to protein+ligand and water+ions sub-systems (2 groups) and to the whole system (one group) were evaluated. The simulation of Langevin dynamics with GROMACS is about 10-15 % slower than MD using either of the thermostats.

The relative residence times for these compounds obtained from τRAMD simulations using the NAMD[29] software with standard parameters (i.e. Langevin thermostat with a relaxation period of 1ps$^{-1}$) correlate well with the measured values (**Fig. 3**). For comparison, we performed GROMACS simulations using several conditions: Langevin dynamics and no thermostat, or Nosé–Hoover or Berendsen thermostat and a Parrinello-Rahman barostat. In the latter two cases, the same relaxation time parameter of 1 ps$^{-1}$ was employed, whereas in Langevin dynamics, the relaxation time parameter was doubled to 2 ps$^{-1}$ as suggested in the GROMACS manual. Simulations with GROMACS under Langevin dynamics conditions with a relaxation time parameter of 1 ps$^{-1}$ yielded approximately twice as long residence times for all compounds. Since simulation with Langevin dynamics is about 10-15% slower than with a Nosé–Hoover thermostat, we used the latter in all subsequent simulations in this study. We also tested using different thermostat groups (either complete system or ions with solvent separated from protein and ligand), but did not notice any significant difference. In summary, for RAMD simulations with GROMACS under all tested conditions, the computed $\tau_{RAMD}$ is very similar to that obtained with NAMD simulations with only small deviations within the computational uncertainty (see **Fig. 3** and **Fig. S1**).

The performance of the GROMACS RAMD simulations on a GPU node is about 3-7 times and on a CPU cluster is more than 10 times better than for the NAMD CPU implementation, giving about 146ns/day and 327 ns/day, respectively, for the HSP90 system (**Table 2**). This difference is achieved because the limitation of serial RAMD steps is overcome and the performance for RAMD is comparable to that for conventional MD in the GROMACS GPU implementation. The new implementation also makes the simulation of larger systems with RAMD computationally feasible. For the M2 receptor system, 34 ns/day was achieved for both conventional MD and RAMD.

*Table 2* Performance of conventional MD and the RAMD procedure as implemented in Gromacs and in a tcl wrapper for NAMD on a CPU-based architecture. The number of cores given corresponds to the best performance (see complete performance plot in **Fig. S3**). Note that the scaling of NAMD workflow is limited by the external tcl script controlling the RAMD simulation procedure. Times are given for simulations of two protein-ligand systems: HSP90 and the M2 muscarinic receptor bound to compound 20 and iperoxo, respectively.

| System | Number of atoms | Conventional MD | | | | | | RAMD | | | | | |
|---|---|---|---|---|---|---|---|---|---|---|---|---|---|
| | | NAMD 12 (MPI+OMP)[a] | | Gromacs 2020 (MPI+OMP)[a] | | Gromacs 2020 single GPU[b] | | NAMD 12 (MPI+OMP)[a] | | Gromacs 2020 (MPI+OMP)[a] | | Gromacs 2020 single GPU[b] | |
| | | cores | Performance (ns/day) | cores | Performance (ns/day) | | Performance (ns/day) | cores | Performance (ns/day) | cores | Performance (ns/day) | | Performance (ns/day) |
| HSP90 | 27 000 | 384 | 138 | 384 | 476 | | 225 | 240 | 20 | 384 | 327 | | 146 |
| M2 | 120 000 | 384 | 36 | 384 | 147 | | 44 | 240 | 12 | 384 | 158 | | 38 |

[a] Intel Xeon E5-2630v3; 4 OMP threads per MPI threat used for Gromacs simulations

[b] Intel Xeon Gold 5118 with NVIDIA Tesla P40; running on 1 node, 4 CPU cores and 1 GPU

**B. Benchmark of the IFP generation protocol**

We first compared computed protein-ligand IFPs for the three HSP90-ligand complexes with those obtained previously[36] using the OpenEye OEChem Toolkit[34] and those generated by the ligand interaction tool of the RCSB PDB[49] and PLIP[42] (see **Table 3**). The number of hydrophobic contacts, HY, in the present study is generally smaller than in Ref. [36] due to the stricter definition of the hydrophobic atoms (i.e. not all carbon atoms are considered as hydrophobic). Apart from HY, there are only a few differences in the IFPs detected by the different methods. For example, the definition of aromatic interactions is less strict in the present study compared to the OpenEye OEChem Toolkit, whereas it agrees well with the ligand interactions identified in the RCSB PDB and PLIP. For h-bond (HD/HA) contacts, differences are observed for compounds 8 and 16, where interactions with T184 are missing due to the distances being slightly longer than the H-bond detection threshold. The computations of water bridges, WB, between the protein and the ligand are found to show the most deviations between methods, with several contacts missed by MD-IFP.

We next benchmarked the IFP detection procedure on 40 complexes from Ref. [43] by comparing the MD-IFP results with those of four programs: PLIP[42], FLIP[43], LPC[56], and MOE[57] (the results are summarized in the **Supplementary EXCEL Table**). Among the 250 PL interactions identified by MD-IFP (excluding hydrophobic interactions), 5 were classified as false positives (two hydrogen bonds and three aromatic interactions), i.e. they were not found by any of the four methods used for the benchmark. One of the hydrogen bonds is a weak hydrogen bond with a fluorine atom considered as an acceptor in PDB ID:3SHY (although C-F⋯H-X is weak, it was shown to be relevant for ligand-protein binding[58]). Remarkably, there are only three false positives amongst the 56 detected aromatic interactions, which indicates that using solely the distance criterion for $\pi$-$\pi$ interactions is sufficient in the majority of cases. Furthermore, the angle-dependent interactions of protein residues with halogen atoms are all correctly recognized by MD-IFP.

Six further interactions (five hydrogen bonds and one salt bridge) were not recognized by MD-IFP and classified as false negatives as they were identified by all of the other methods.

*Table 3 Comparison of the IFPs computed for the crystal structures of three HSP90-inhibitor complexes with MD-IFP, in Ref. [37] using the OpenEye OEChem Toolkit[34], from the ligand interaction diagram in the RCSB PDB database[49], and with PLIP[42]. a)*

| PDB ID | Ligand | MD-IFP | Ref. [37] (using OEChem) | Ligand interactions in the RCSB PDB database[49] b) | PLIP[42] |
|---|---|---|---|---|---|
| 5J64 | 8 | *AR*: K58<br>*HD/HA*: N51 K58 D93 GLY97<br>*WB*: L48 S52 | *HD/HA*: D93 G97 **T184** | *AR*: K58<br>*HD/HA*: K58 D93 G97<br>*WB*: L48 S52 **T184 G97** | *AR*: K58<br>*HD/HA*: K**51** D93 G97 **T184**<br>*WB*: L48 **N51 G95** |
|  |  | *HY*: N51 M98 T184 | *HY*: N51 **S52 D54 A55** I96 GL97 M98 **L107** G108 T109 **F138** T184 **V186** |  | *HY*: T184 |
| 5J86 | 16 | *AR*: K58 F138 W162<br>*HD/HA*: N51 K58 D93 G97<br>*WB*: L48 <u>S52</u> | *HD/HA*: N51 D93 G97 | *AR*: N51 K58 W162<br>*HD/HA*: K58 D93 G97 **T184** | *AR*: K58 F138 W162<br>*HD/HA*: N51 D93 G97 **T184**<br>*WB*: L48 **G95** |
|  |  | *HY*: N51 D54 M98 L103 L107 W162 T184 | *HY*: N51 **S52** D54 **A55 D93** I96 G97 M98 **L103** L107 **F138 L150 W162** T184 **V186** |  | *HY*: D54 L103 T184 |
| 5LQ9 | 20 | *AR*: F138 W162 F170<br>*HD/HA*: K58 Y139 184 S52 D93<br>*WB*: K58 L48 | *AR*: F138 W162 F170<br>*HD/HA*: D93, Y139 | *AR*: F138, W162<br>*HD/HA*: K58 D93, Y139<br>*WB*: L48 **S52 D93 G97 T184** | *AR*: F138 W162<br>*HD/HA*: K58 D93, **T184**<br>*WB*: **S52**, K58 |
|  |  | *HY*: F22 M98 L103 L107 F138 V150 W162 F170 | *HY*: F22 **Q23 N51 S52 D54 A55 A57 K58 D93** I96 **G97** M98 L103 L107 **G108 I110 A111** F138 **Y139** V150 W162 F170 **T184 V186** |  | *HY*: F22 L103 L107 Y138 V150 Y170 |

a) IFPs that are not identified by MD-IFP are shown in bold and those that were not identified by any of the other methods used for benchmarking are underscored.

b) In the RCSB PDB database[49], hydrophobic interactions are not included.

As expected, the major inconsistency between the different methods comes from the WB detection. In the present work, 69 WBs were identified. Among them, 23 were not found by the benchmark methods (FLIP, PLIP, MOE), which indicates that criteria of WB detections are less strict defined as in the present method. However, only four detected WBs were found by all the latter methods but were missing in MD-IFP. Inconsistency in the identification of water bridges partially arises from differences in the approaches used and in the hydrogen bond parameters employed, as well the ambiguity of the assignment of hydrogen orientation (which depends on the procedure used for protonation). Interestingly, almost half of the false positives correspond to plausible water bridges between the protein backbone and the ligand (some examples are illustrated in **Figure S4**)

**C. Analysis of ligand dissociation routes**

The workflow developed here includes tools to analyse ligand dissociation routes on the basis of clustering in IFP space and a network analysis of the clusters. Since ligands spend the most of the RAMD simulation time in its bound state, we extracted for analysis the last 300 frames (i.e. the last 300ps) from each dissociation trajectory (see ligand RMSD variation in **Fig. S5**, and more details in Appendices, section E. Analysis Protocols). We show the capabilities of these tools by analysing the three inhibitors of HSP90 illustrated in **Fig.2**. These inhibitors differ in size and have quite distinct dissociation pathways: compounds 8 and 16 are relatively small and occupy only the ATP binding site, whereas compound 20 has a quinoline fragment that occupies the hydrophobic subpocket located under α-helix3 (see **Fig.2C**). The dissociation pathways in the IFP space for the three compounds are represented as a network between clusters in **Fig.4** (from k-means clustering with 8 clusters, see Appendix E for details). Nodes denote clusters with their size proportional to the cluster population and the color density increases with the average ligand RMSD in the cluster from the initial bound position. The clusters are ordered along the x-axis by the average displacement of the ligand COM in the cluster from the starting structure. Ligand motion is shown as transitions between clusters, which are also used to compute and visualize the net flow between nodes (see **Fig.4**).

The bound states can be distinguished from intermediate ones by the small value of the ligand RMSD relative to the starting ligand position (within 2 Å; bound states are indicated by light-orange circles, accordingly). Usually, the RMSD from the starting structure is small and there are only slight variations in IFPs for all bound-state clusters. Note, however, that we here analyze only the last 300 snapshots and even in the first snapshot analyzed, the ligand may have a slightly different position from in the bound state from which RAMD simulations were started.

There is one dissociated state (without or with no specific PL contacts) that has a large RMSD value (> 10Å) and is colored dark-orange. The rest of the nodes can be considered to be metastable states along the ligand dissociation pathway. The number of metastable states naturally depends on the complexity of the egress route. Specifically, for the smallest compound, 8, only one intermediate state (cluster 7, **Fig. 4A**) is identified, which is very close to the bound states and differs from them by the loss of the interactions with GLY97, MET98, and THR184 (see **Fig 4B**). This metastable state is shown in **Fig. 5A** by the COM distribution of the cluster members mapped on to a 3D grid. Although less pronounced, direct dissociation from the bound state (e. g. cluster 6) is also observed. All dissociation routes lead directly from the ATP binding pocket (**Fig. 5B**).

For compound 16, there is also only one intermediate state (cluster 7, **Fig.4C, Fig.5C**), but it is located further from the bound states on the COM scale (about 5 Å). In the IFP profile, only contacts to MET98, LEU107 and PHE138 and a water bridge to ASP93 are preserved, while the other interactions are much less pronounced (**Fig. 4D**). Unlike compound 8, where the hydrogen bond with ASP93, the main anchor point for all compounds bound to the ATP binding site of HSP90, is lost only upon complete dissociation (i.e. in the metastable state 8), for compound 16 this hydrogen bond is first broken upon transition to the metastable state 7, where the compound still retains multiple hydrophobic and hydrogen bonds and gains a new water-mediated contact to ASN51.

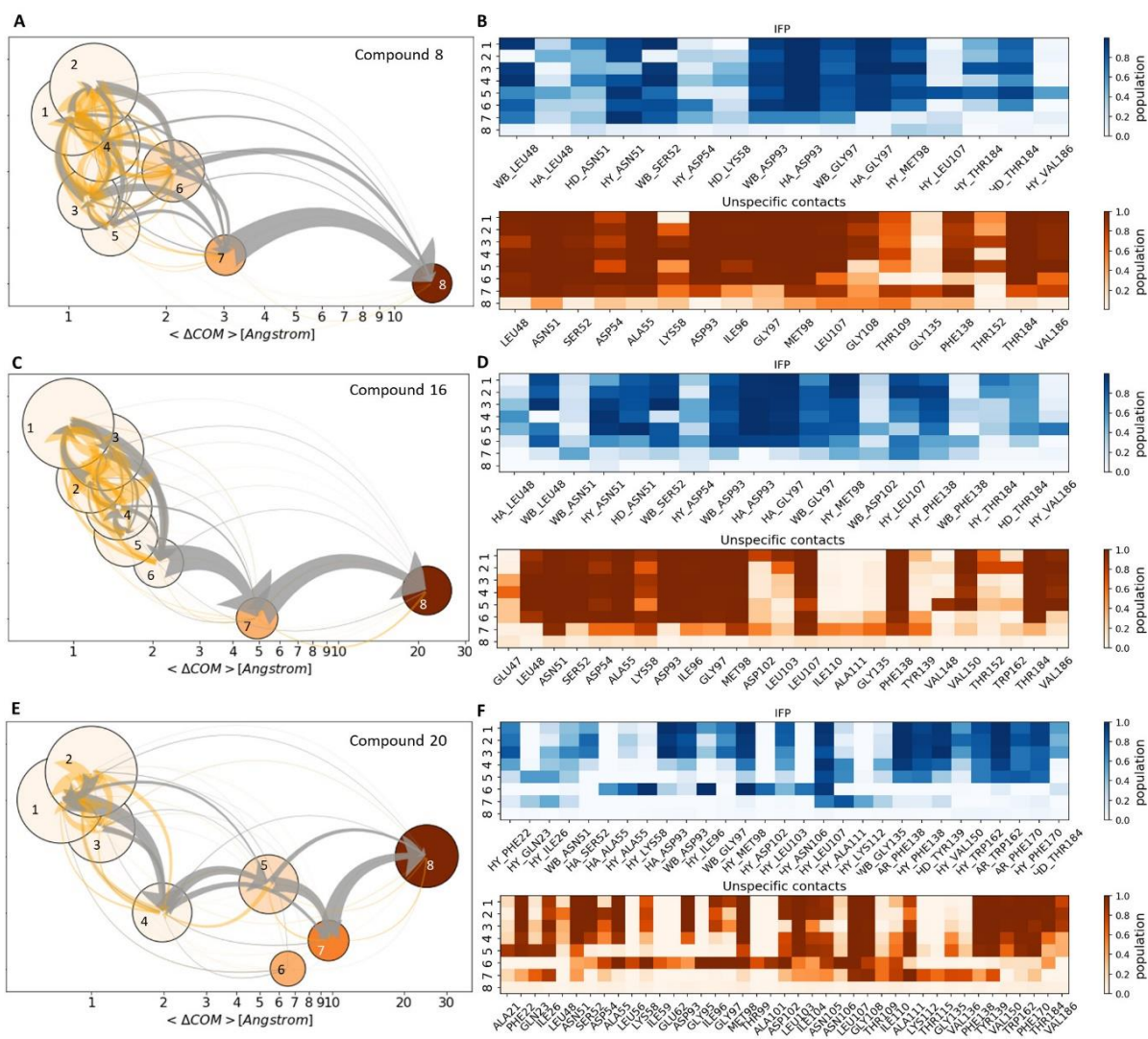

***Figure 4***: *Schematic visualization of the RAMD dissociation trajectories (the last 300 frames of each trajectory are considered) of 3 inhibitors of HSP90, compounds 8 (A,B), 16 (C,D), and 20 (E,F): (A, C, E) Dissociation pathways are shown in a graph representation. Each cluster is shown by a node with the size indicating the cluster population. Nodes $C_i$ are positioned on an increasing logarithmic scale of the average ligand COM displacement in the cluster from the starting snapshot and the node color denotes the averaged ligand RMSD in the cluster from the starting structure. The width of the light-orange arrows is proportional to the number of corresponding transitions ($C_i$ -> $C_j$) and ($C_i$ <- $C_j$) between two nodes $C_i$ and $C_j$ and the gray arrows indicate the total flow between two nodes (i.e. transitions ($C_i$ -> $C_j$)-($C_i$ <- $C_j$)). (B, D, F) - IFP composition of each cluster. PL IFPs and nonspecific protein-ligand contacts within a distance threshold of 5Å between heavy atoms (in blue and orange pallets, respectively). A 2D Euclidian distance matrix in the IFP space between cluster means is shown in Fig. S6.*

The structure of the egress paths becomes more complicated for the bulkier and more slowly dissociating compound 20, which passes through multiple intermediate transient states during dissociation. In contrast to the smaller compounds, which demonstrate a single dissociation route, compound 20 has two possible egress routes (indicated by the red arrows in **Fig. 5F**). One route goes directly from the ATP-binding site through the intermediate states 6 or 5 (gray and yellow isosurface in the COM distribution in **Fig. 5E**). The other route runs through the transient hydrophobic subpocket under α-helix3 (via clusters 5 and 7, (**Fig. 5E**).

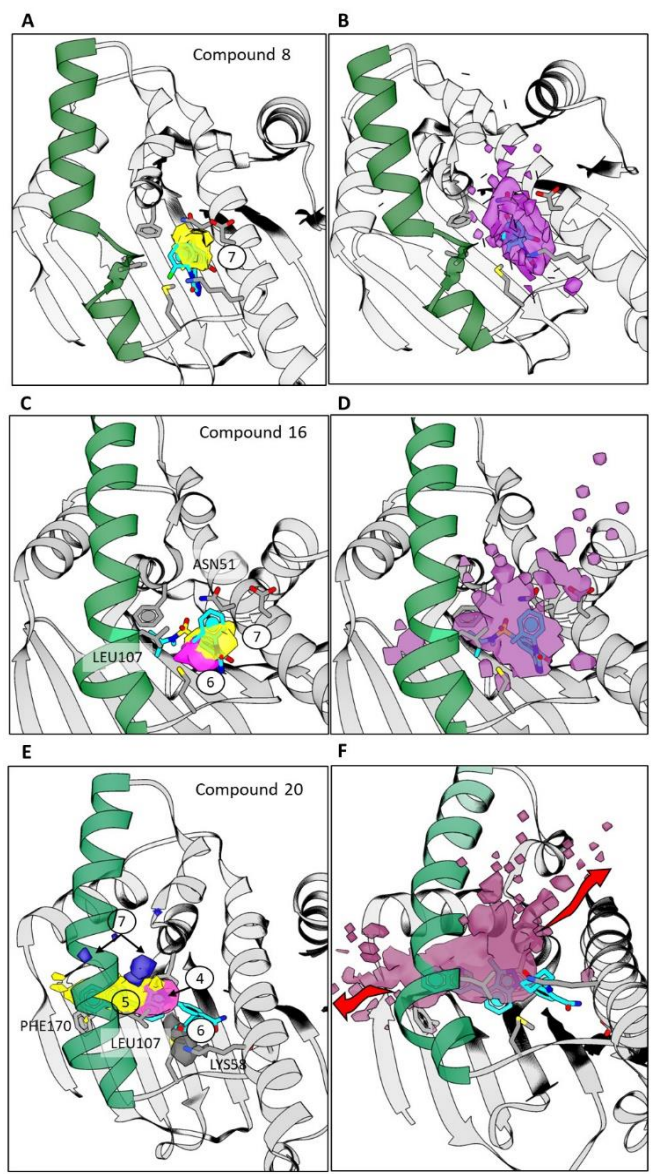

*Figure 5: RAMD dissociation trajectories for 3 inhibitors of HSP90. Compounds 8 (A,B), 16 (C,D), and 20(E,F) are shown with carbons in cyan and with α-helix3 in green. (A, C, E) – The positions of the main metastable states are shown by iso-surfaces of different colors labelled by the corresponding cluster number in Fig.4. (B, D, F) – Ligand COM population density from the last 300 frames of each trajectory shown by iso-surfaces clipped in the visualization plane. For compound 20, two egress routes are indicated by red arrows.*

The IFP composition of the clusters in **Fig. 4E** shows that compound 20 loses its hydrogen bond to ASP93 in all metastable states starting from the cluster 4 and instead forms new contacts with α-helix3 (such as LEU107, ILE110, ALA111). These contacts are all rather nonspecific (e.g. hydrophobic), which indicates that the long dissociation times for the pathway under α-helix3 are mainly due to limited space in the dissociation tunnel rather than the formation of specific interactions.

### III.     CONCLUSIONS

In this paper, we present an efficient implementation of the RAMD method along with analysis tools for the exploration of ligand egress routes and for the prediction of relative protein-ligand residence times using the τRAMD approach[36,37]. Despite the promising efficiency and accuracy of the method, the preceding implementation suffered from two important limitations: (i) the parallel scaling performance of MD simulations was diminished as the random force adjustment steps were serially wrapped around the MD engine, and (ii) difficulties in the quantitative analysis of large numbers of ligand dissociation trajectories, which hindered the detection of possible egress routes and mechanisms, and their statistical analysis. To overcome these two bottlenecks, we developed a new open-source workflow that includes: (i) a new implementation of the RAMD method in the GROMACS MD simulation package that significantly improves simulation performance relative to the previous implementation in NAMD, and (ii) a Python-based workflow for automated analysis of ligand egress trajectories, which includes computation of residence times for series of compounds, and the MD-IFP tool set for generation of protein-ligand interaction fingerprints along ligand dissociation trajectories, and analysis of ligand dissociation pathways in the IFP space.

The new implementation of the RAMD protocol in the GROMACS PULL function speeds up simulations of ligand dissociation by more than ten times relative to the previous implementation in a tcl script with the NAMD engine. We evaluated the new implementation for complexes of three inhibitors with HSP90, a system previously studied in detail[36], using different parameters for the underlying MD protocol. We found that the results of the τRAMD procedure obtained using the two MD simulation programs are the same within the uncertainty of simulations. Then we compared the performance of the two implementations on systems of different sizes. These simulations showed that the sampling of ligand egress for computing a relative residence time (usually about 60 ligand dissociation trajectories up to several nanoseconds duration were required) could be performed within about half a day on one GPU node for a relatively small system (the solvated globular HSP90 domain with 27 500 atoms) and within two days for a larger system (a GPCR protein embedded in a lipid bilayer having about 120 000 atoms). Taking into account that the individual RAMD simulations can be performed independently and, thus effectively parallelized, the method provides the possibility to compute relative residence times for multiple drug candidates within a few days on a GPU or CPU cluster. Furthermore, we report a protocol for processing the output RAMD trajectories that enables the automated computation of relative residence times, their statistical assessment, and the comparison of computed with experimental data, if available.

Analysis of the dissociation trajectories is another important aspect of the new workflow. The τRAMD method is aimed at the quick estimation of the dissociation rates and requires tens of trajectories to be generated per ligand to ensure reasonably small uncertainty in the computed values. Nonetheless, even hundreds of generated trajectories cannot completely cover the full configuration space of the multidimensional ligand-protein dissociation landscape. For this reason, reconstruction of the protein-ligand dissociation free energy landscape or even the free energy profile of a single dissociation pathway, is not feasible. One should, however, expect that the generated trajectories bear important information about protein-ligand interactions that affect residence time if the derived residence times provide the correct trends for a set of ligands. Therefore, we developed MD-IFP to generate protein-ligand IFPs for MD trajectories. We benchmarked MD-IFP on 40 protein-ligand complexes and found that it identified similar IFPs to several available methods. Then we applied the procedure to trajectories of three complexes of HSP90 with inhibitors that have very different residence times, sizes, and binding poses and, thus, IFP profiles. We showed how the workflow could be used to explore the increasing complexity of the dissociation pathway in the IFP space on increasing ligand size, which, in this case, is correlated with increasing residence time.

In summary, the workflow reported provides an efficient computational engine for the estimation of the relative residence times of compounds against a macromolecular target and tools for obtaining insights into the underlying mechanisms determining ligand unbinding kinetics. It may thus facilitate the assessment and selection of drug candidates in the early stages of a drug development pipeline.

## SUPPLEMENTARY MATERIAL

Figures S1-S9 provide additional results:

Figures S1: RAMD simulations using the GROMACS implementation for three HSP90 compounds showing a comparison of using the Langevin dynamics, Berendsen, and Nose-Hoover thermostats; Figure S2: Illustration of the procedure for detection $\pi$-$\pi$ interactions with MD-IFP; Figure S3: Plot showing performance of different implementations for two test examples: HSP90 and Muscarinic receptor M2; Figure S4: Illustration of water bridges identified by MD-IFP; Figure S5: Structural variations in the last 300 frames of RAMD dissociation trajectories for three compounds; Figure S6: 2D Euclidian distance matrix in the IFP space between cluster means for the clusters; Figure S7: Illustration of the effect of cluster number on the structure of the simulated dissociation pathways; Figure S8: Illustration of the clustering obtained by applying the Gaussian Mixture method to the RAMD dissociation trajectories of the 3 compounds; Figure S9: Illustration of dissociation networks generated for the systems shown in Figure S7 using clustering based on the IFP only.

Excel table containing results of the benchmark of the MD-IFP protocol for 40 protein-ligand complexes.

## DATA AVAILABILITY STATEMENT

GROMACS-RAMD version 1.0 has been released for GROMACS versions 2019 and 2020 and is publicly available at https://github.com/HITS-MCM/GROMACS-ramd

Tutorials for the τRAMD protocol implemented with NAMD and GROMACS are available on KBbox: https://kbbox.h-its.org/toolbox/

Python scripts of the IFP generation and analysis are available at: https://github.com/HITS-MCM/MD-IFP

The data that supports the findings of this study are available within the article and on reasonable request to the authors.

**AUTHORS' CONTRIBUTIONS**

DBK and RCW conceived the research; BD implemented the RAMD procedure in the GROMACS software; SR assisted in software design and implementation; DBK performed computations and DBK, FO and XC analyzed the computational data; FO carried out benchmark of the IFP computation protocol; DBK and RCW wrote the manuscript with contributions from BD and FO.

**ACKNOWLEDGEMENTS**

This research work has received funding from the European Union's Horizon 2020 Framework Programme for Research and Innovation under the Specific Grant Agreement No. 785907 and No. 945539 (Human Brain Project SGA2 and SGA3). We also thank the Klaus Tschira Foundation for support. We are grateful Ariane Nunes-Alves for testing IFP scripts and for useful suggesting regarding the manuscript.

**APPENDICES**

**METHODOLOGY/IMPLEMENTATION/TECHNICAL DETAILS**

**A. tcl implementation of RAMD in NAMD**

In the NAMD-based protocol for RAMD simulations, a tcl script [16,19] wrapped around the MD is employed. The script computes the ligand displacement after each short (50 timesteps) MD interval and recomputes the force if necessary. It sends a kill signal to stop the trajectory when the ligand displacement from its initial position reaches the predefined dissociation threshold distance. In the recent version the procedure of selection force orientation was improved to ensures uniform distribution of the vector direction (version 5.05 https://www.h-its.org/downloads/ramd/ )

**B. Implementation of RAMD in GROMACS**

The RAMD implementation is based on the 'pull' code in GROMACS. The key feature of the pull code is that forces are applied between the centers of mass of pairs of atom groups.

Core functionalities like the usage of MPI and/or GPU are already included and the performance has been optimized. In order to keep the interface as user-friendly as possible, only RAMD settings have to be provided. All pull code-related settings are handled automatically by the RAMD implementation during the GROMACS preprocessing step.

A function for changing the pull direction during the simulation is not available in the pull code. Therefore, the random force direction is decomposed into three orthogonal unit vectors (1,0,0), (0,1,0) and (0,0,1) and only the projected force values have to be adjusted. For testing that the force directions were distributed in a spherically uniform fashion, a sphere was divided into 32 longitudinal bins and a large number (1 billion) of force directions was generated randomly. The procedure was repeated in each direction in space to ensure a uniform spherical distribution. By counting the number of force vectors assigned to the bins, we could ensure that the force generator indeed covers the whole sphere uniformly.

**C. Validation of MD-IFP on a set of crystal structures of protein-ligand complexes**

For the validation, we chose the same set of structures of protein-ligand complexes that was used to validate FLIP[33]. It consists of 50 RCSB PDB entries and their corresponding ligand and chain identifiers. Hydrogens were added using Chimera[45] 1.14, using the "unspecified" protonation state with the consideration of hydrogen bonds. A protein chain, a ligand and water molecules were extracted using the pdb-file processing tools in the Biopython (version 1.76) package[59]. For structures containing multiple conformations, conformation A was extracted and all other conformations were discarded. The ligand was saved in mol2 format. For structures containing azole, amidine and urea groups, Chimera failed to produce correct mol2 files and therefore, these files were manually corrected. If there was no apparent reason for the mol2 file not working, we created the mol2 file using MOE[57].

Altogether 10 structures were removed from the original dataset because we were either unable to generate usable mol2 files or the structures were unsuitable for our processing pipeline (e.g. having two ligands covalently bound together), or other methods using for benchmark were unable to generate results. Thus, our final benchmark set consisted of 40 structures.

The IFPs generated by MD-IFP were compared to the results from FLIP[33], PLIP[32], LPC[46], and MOE[57]. We considered the interactions detected by MD-IFP as False Positives (FP) if they could not be detected by any of these four methods and as False Negatives (FN) if all four methods detected them but they were not found by MD-IFP. Since water bridges and halogen bonds could not be detected using LPC, we classified them as FN if FLIP, PLIP, and MOE detected the interaction.

Hydrophobic interactions were not considered, since their definition is not as clear cut as for the other interactions and one would naturally expect a lot of variation between methods for them. Except for MOE, none of the benchmark methods differentiate between donated and accepted hydrogen bonds, both indicated by the abbreviation HB for hydrogen bond in the "missing interaction" column. Since we consider fluorine to make hydrophobic interactions, halogen bonds with fluorine were excluded from the analysis. Interactions with metal ions and cofactors were excluded, since the ability to detect them has not yet been implemented in MD-IFP.

Benchmark results are summarized in the **Supplemented excel table**.

### D. Details of MD simulations

**System setup and force field parameters**

The structures of the HSP90-inhibitor complexes were prepared from the crystal structures with PDB ID: 5J64, 5J86, 5LQ9 for compounds 8, 16, and 20, respectively, as described in Ref.[36]. The structure of the M2 muscarinic GPCR with the orthosteric ligand, iperoxo, bound was prepared from the structure with PDB ID 4MQT[60]0

 with the allosteric compound LY2119620 removed. The CharmmGUI[61] web server was used to embed the GPCR in a pure 1-palmitoyl-2-oleoylphosphatidylcholine (POPC) bilayer and perform protein protonation, generation of topology files and coordinates for AMBER simulations. Systems were solvated with TIP3P[62] water molecules with a margin of at least 10 Å from the protein and $Na^+$ and $Cl^-$ ions were added to ensure system neutrality at an ion concentration of 150mM. Iperoxo was modelled in its protonated state (charge +1e). For all systems, the Amber ff14[63] and GAFF[64] force fields for protein/lipid and ligands, respectively, were employed. RESP partial atomic charges for ligands were computed using GAMESS[65] calculations of the electron density population at the HF/6-31G*(1D) level and Amber tools[63].

**Simulation protocol**

In all cases, the system was first energy minimized and equilibrated using the Amber18 software[66]. For HSP90, a step-wise minimization, heating and equilibration was done as described elsewhere[36]. The system with the membrane protein was first minimized (restraints on all heavy atoms except water and ions of 1000, 500, 100, 50, 10, 1, 0.5, 0.1, 0.05, 0.01 kcal $mol^{-1}$ $Å^{-2}$ for 1000 steps of conjugate gradient and then 10000 steps without restraints), then heated in 200 ps steps with restraints of 100 kcal $mol^{-1}$ $Å^{-2}$ on all heavy atoms except water and ions up to 100 K (NVT- Langevin tau = 1 $ps^{-1}$) and then in an NPT ensemble (7 ns) up to 310K with decreasing restraints of 50, 30, 5 kcal $mol^{-1}$ $Å^{-2}$ and finally without restrains. Then we followed the protocol for the setup of simulations of membrane-containing systems on GPUs (https://ambermd.org/tutorials/advanced/tutorial16/) that consists of 10 consecutive simulations of 5ns duration (which is required because the GPU code does not recalculate the non-bonded list cells during a simulation). Finally, we ran a further simulation of 300 ns under NPT (Langevin thermostat with a Berendsen barostat) conditions to ensure equilibration of the whole system. For all simulations, a cutoff of 10 Å for nonbonded Coulombic and Lennard-Jones interactions and periodic boundary conditions with a Particle Mesh Ewald treatment of long-range Coulombic interactions were used. A 2 fs time step was employed with bonds to hydrogen atoms constrained using the SHAKE algorithm[67].

The equilibrated systems were then used in the NAMD and GROMACS tauRAMD protocols. The protocol employed for RAMD simulations using the NAMD[29] software was reported elsewhere[32] and can be found online at (kbbox.h-its.org).

To perform simulations in GROMACS[68], the final output coordinate and topology files were transferred from Amber to GROMACS using ParmEd[69]. Then we first performed short NVT simulations (Berendsen thermostat, 30 ns) and then generated four trajectories under NPT conditions (Nosé–Hoover thermostat and Parrinello-Rahman barostat, 30ns). Each trajectory was started with velocities generated from the Maxwell distribution to ensure trajectory diversity.

RAMD simulations were performed with GROMACS at NPT conditions (Nosé–Hoover thermostat and Parrinello-Rahman barostat) except for the cases where different thermostats were evaluated. Displacement of the ligand COM was checked every 100 fs and then the random force orientation was either retained (if the ligand COM had moved by at least 0.025 Å) or changed randomly otherwise. Simulations were stopped when the ligand COM had moved further than 30 Å from protein COM in HSP90 and further than 50 Å in the M2 receptor. Coordinates were saved at 1 ps intervals.

### E. Analysis Protocols

**Preprocessing**

IFPs were generated for the 300 last frames of each trajectory (superimposed with the last snapshot of the equilibration trajectory employed as a reference), thus discarding the majority of the frames where the ligand retains a bound state position. The IFPs for all frames, together with nonspecific contacts within a threshold distance of 5 Å, were collected in one binary matrix for each compound filled with 0/1 values for each particular contact (i.e. residue and type of interaction) and frame. Additionally, RMSD of the ligand and protein relative to reference as well as the ligand COM coordinates and the number of water molecules in the ligand solvation shell were stored.

**Clustering**

We employed k++-means clustering as implemented in the scikit-learn package[70] to detect the most visited regions in the IFP space. The clustering was done on the set of IFP vectors and unspecific contacts.

The selection of the number of clusters to be generated is the main bottleneck in the k-means approach. In the present case, we chose the number of clusters from a trade-off between the difficulty in analyzing multiple clusters and the blurring of the protein-ligand contact specificity in the case of a small number of clusters. Specifically, we selected the minimal number of clusters, 8, for which transient states in dissociation trajectories were clearly recognized for all three compounds (see illustration of dissociation trajectories with smaller and larger numbers of clusters in **Fig. S7**). Note, that a larger number of clusters (i.e. up to 10) does not change essentially the general pattern of dissociation behavior for any of the three ligands, but rather increases the number of clusters that characterize possible configurations of the bound state. On the other hand, if too few clusters are selected (6 or 4), some intermediate metastable states are merged with the bound state. However, even with 4 clusters, the difference in the dissociation profile of compound 20, having well-defined intermediate states, from those of compounds 8 and 16, whose dissociation pathways are reduced to a direct transition from the bound to the unbound state, is clearly apparent.

Additionally, we tested using the Gaussian Mixture (GM) method instead of k-means for clustering (using default parameters of the scikit-learn package). The GM method is based on a probabilistic model that assumes that the data can be described by a mixture of a finite number of Gaussian distributions. The dissociation pathways obtained from GM clustering (see **Fig. S8**) show slightly different distributions of the intermediate metastable states but the general difference in the pattern of states of compounds 8 and 16 versus compound 20 is retained, showing that the latter compound has a notably more complicated dissociation route.

Using just IFP for clustering, instead of a combination of IFP and nonspecific protein-ligand contacts, does not change this pattern either, although the minimum number of clusters required to reveal a metastable state for compound 16 increases from 8 to 10 (**Fig.S9**).

# SUPPLEMENTARY MATERIAL

# A Workflow for Exploring Ligand Dissociation from a Macromolecule: Efficient Random Acceleration Molecular Dynamics Simulation and Interaction Fingerprints Analysis of Ligand Trajectories.

Daria B. Kokh[a*], Bernd Doser[b], Stefan Richter[a], Fabian Ormersbach[a], Xingyi Cheng[a, c], Rebecca C. Wade[a,d,e*]

[a]Molecular and Cellular Modeling Group, Heidelberg Institute for Theoretical Studies, Schloss-Wolfsbrunnenweg 35, 69118 Heidelberg, Germany

[b]Heidelberg Institute for Theoretical Studies, Schloss-Wolfsbrunnenweg 35, 69118 Heidelberg, Germany

[c]Molecular Biosciences, Heidelberg University, Im Neuenheimer Feld 282, 69120, Heidelberg, Germany

[d]Center for Molecular Biology (ZMBH), DKFZ-ZMBH Alliance, Heidelberg University, Im Neuenheimer Feld 282, 69120 Heidelberg, Germany

[e]Interdisciplinary Center for Scientific Computing (IWR), Heidelberg University, Im Neuenheimer Feld 205, Heidelberg, Germany

*Daria.Kokh@h-its.org, Rebecca.Wade@h-its.org

## A  Comp.8

### Langevin dynamics
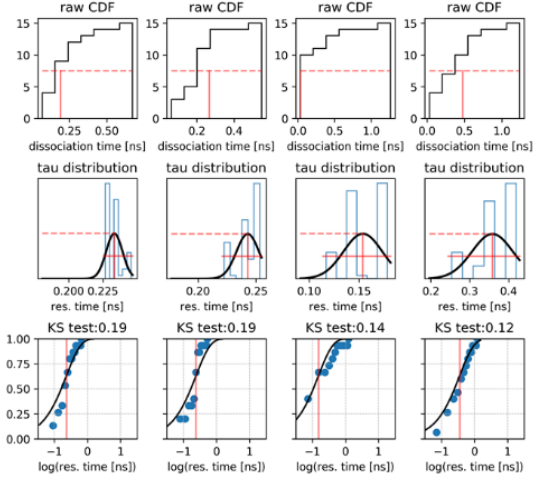
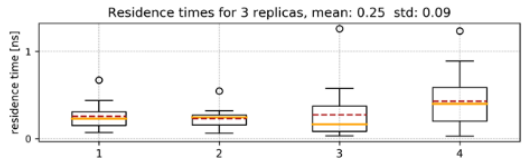

### Berendsen thermostat
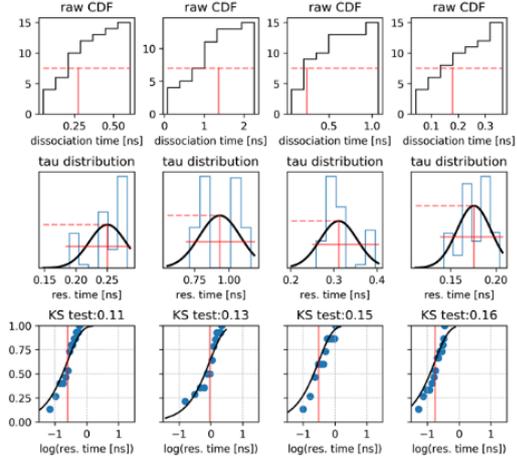
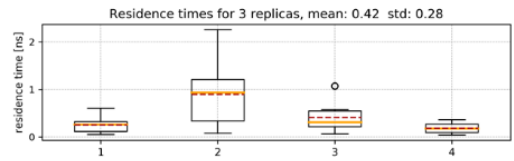

### Nose-Hoover thermostat
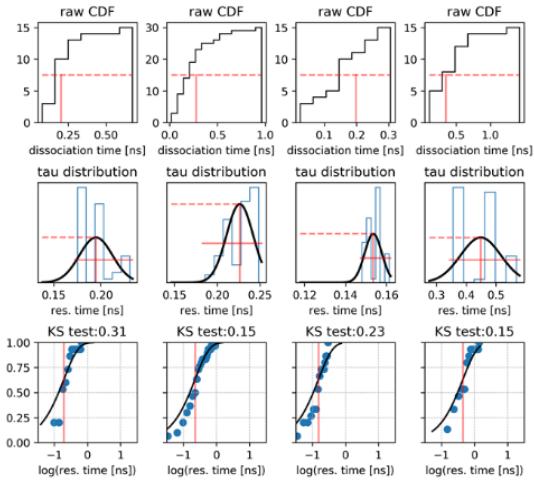
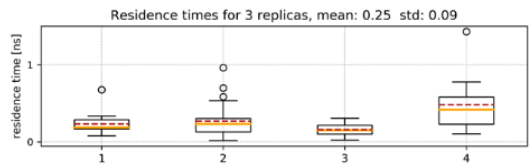

## B Comp. 16

Langevin dynamics

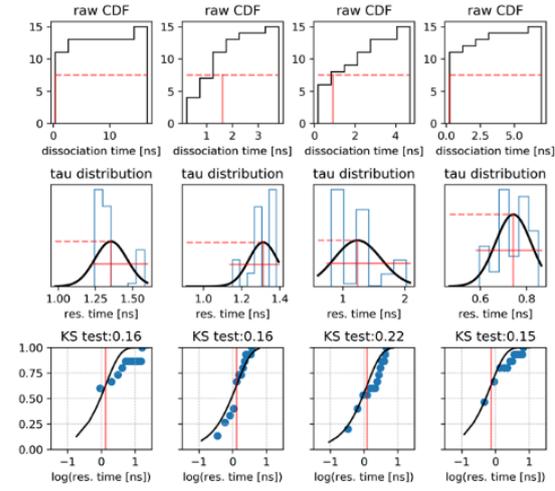

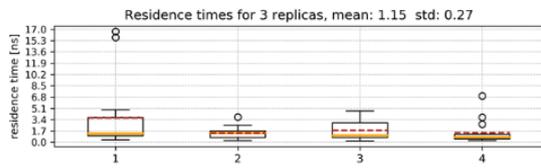

Berendsen thermostat

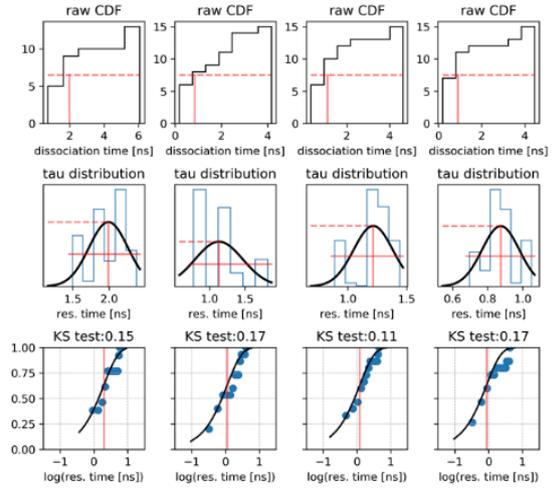

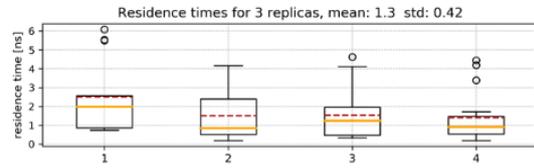

Nose-Hoover thermostat

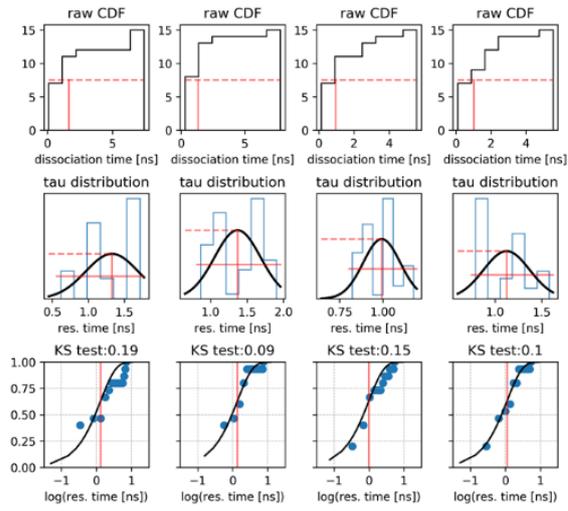

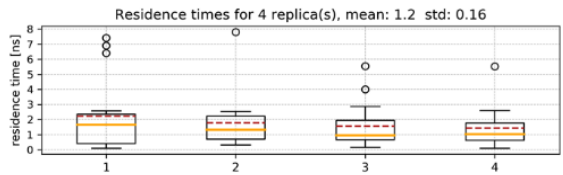

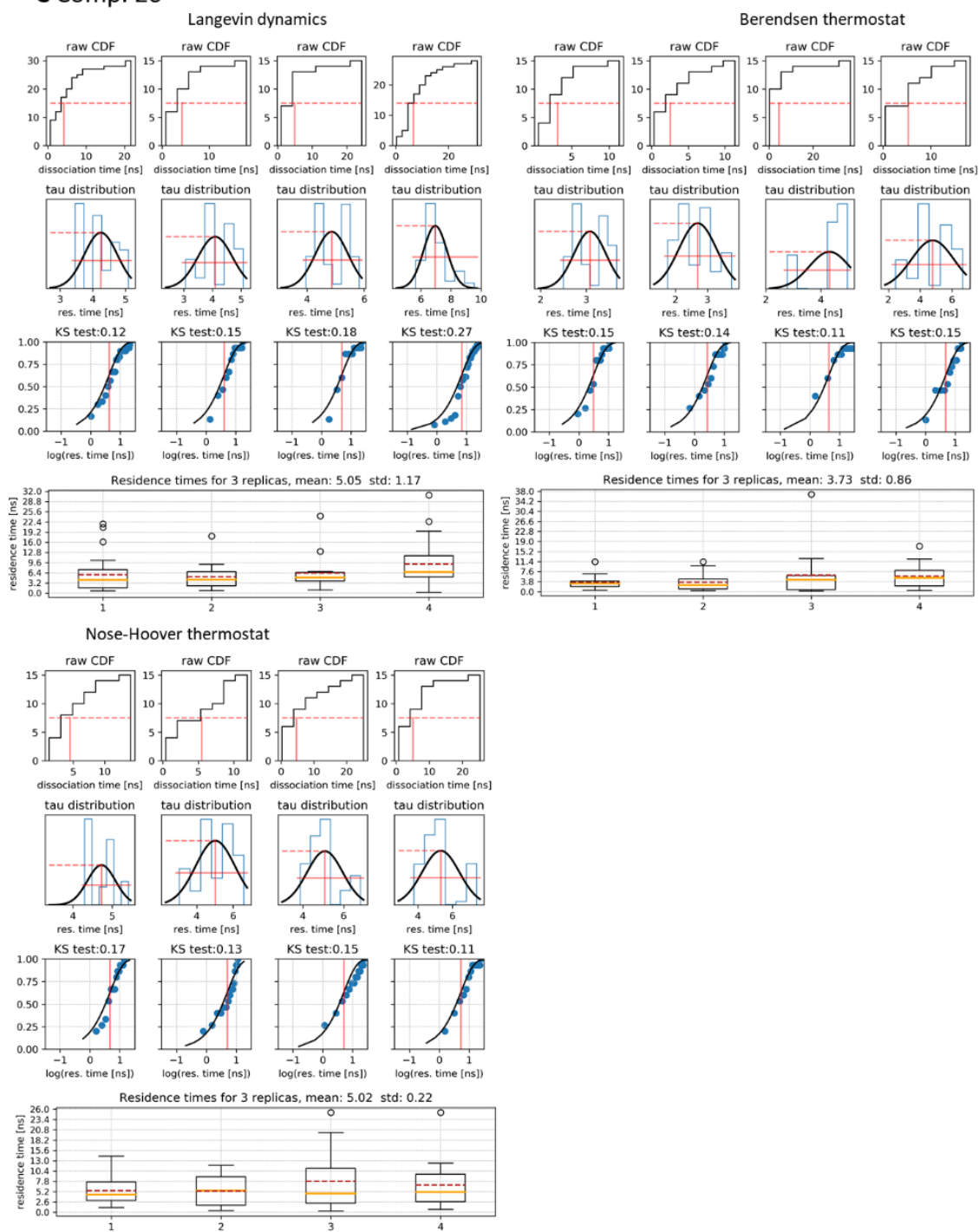

***Figure S1*** *τRAMD simulations using the GROMACS implementation for three HSP90 compounds showing a comparison of using the Langevin dynamics (tau = 2 ps), Berendsen and Nose-Hoover thermostats (tau = 1 ps):* ***(A)*** *Compound 8: the computed residence time averaged over 4 replicas is 0.25±0.09 ns, 0.42±0.28 ns, and 0.25±0.09 ns respectively. For comparison, in NAMD simulations Langevin thermostat with tau = 1 ps), the computed residence time is 0.22±0.06 ns (not shown in the plot);* ***(B)*** *Compound 16: averaged residence time is 1.15±0.27 ns, 1.3±0.42 ns, and 1.2±0.16 ns, respectively; in NAMD simulations the residence time is 1.4±0.29 ns (see* ***Fig. 1*** *of the main text);* ***(C)*** *Compound 20: averaged residence time is 5.05±1.17 ns, 3.73±0.86 ns, and 5.02±1.22ns, respectively; in NAMD simulations the residence time is 4.36±1.8 ns (not shown in the plot).*

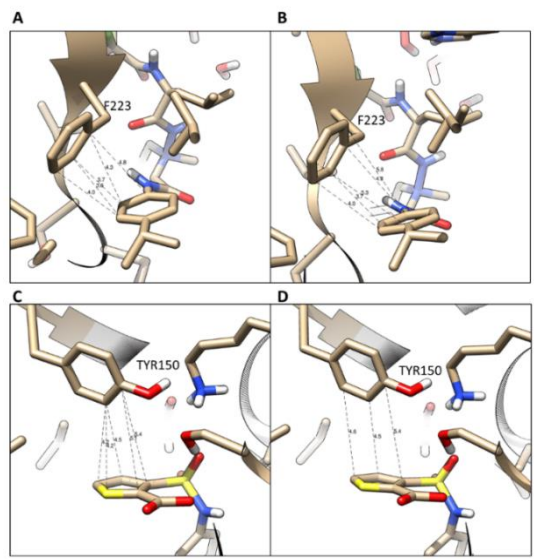

***Figure S2*** *Illustration of detected π-π interactions with MD-IFP for two crystal structures: (A,B) slightly tilted in-plane (PDB ID 1BMA) and (C,D) edge-to-plane (PDB ID 1L2S) interactions. (A,B) In the in-plane interaction, both aromatic rings (the ligand benzene fragment and F223) have more than 5 atoms in the interaction list of atoms within 5.5Å. (C,D) In the edge interaction, the ligand fragment has contacts with 5 (C) or 3 (D) aromatic atoms of TYR150. Both cases were recognized as π-π interactions by MD-IFP and by the ligand interaction tool in the RCSB PDB and by LPC, but not by FLIP or PLIP*

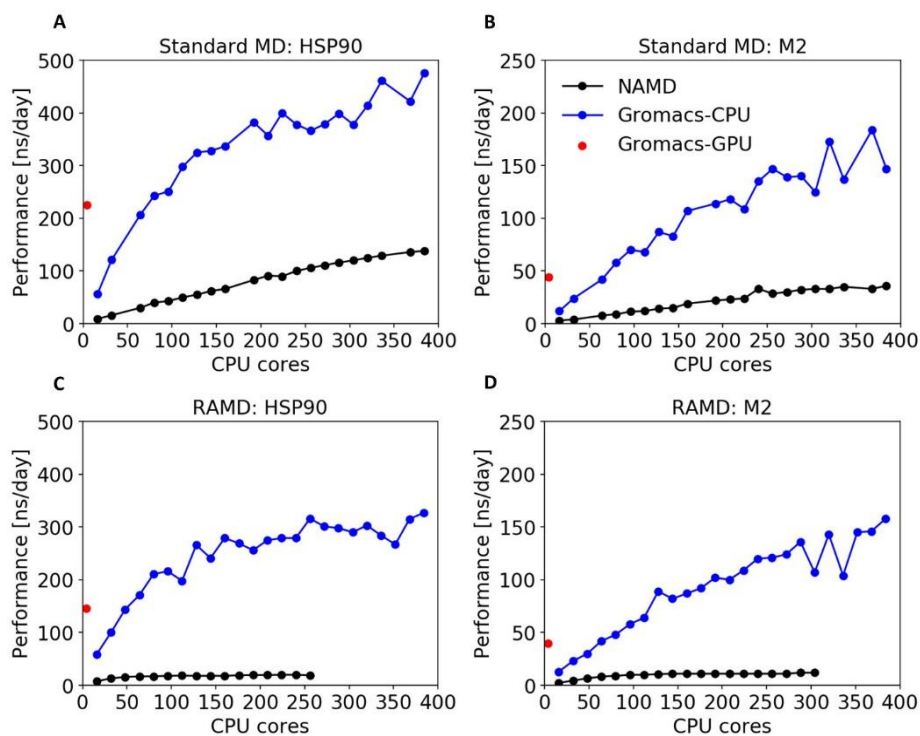

***Figure S3*** *Plot showing performance of two different implementations of the RAMD protocol – in NAMD v.12 and Gromacs 2020 software for two test examples: HSP90 (27 000 atoms) and Muscarinic receptor M2 (120 000 atoms), see more details in Table 2 of the main text.*

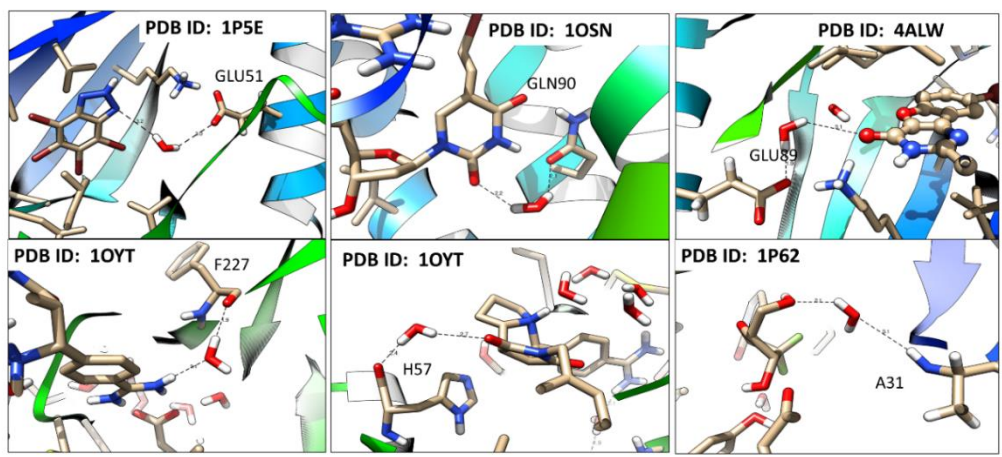

***Figure S4*** *Illustration of water bridges identified by MD-IFP (marked by dashed lines) but not found by any of the benchmark methods and therefore categorized as false positive in the benchmark of MD-IFP.*

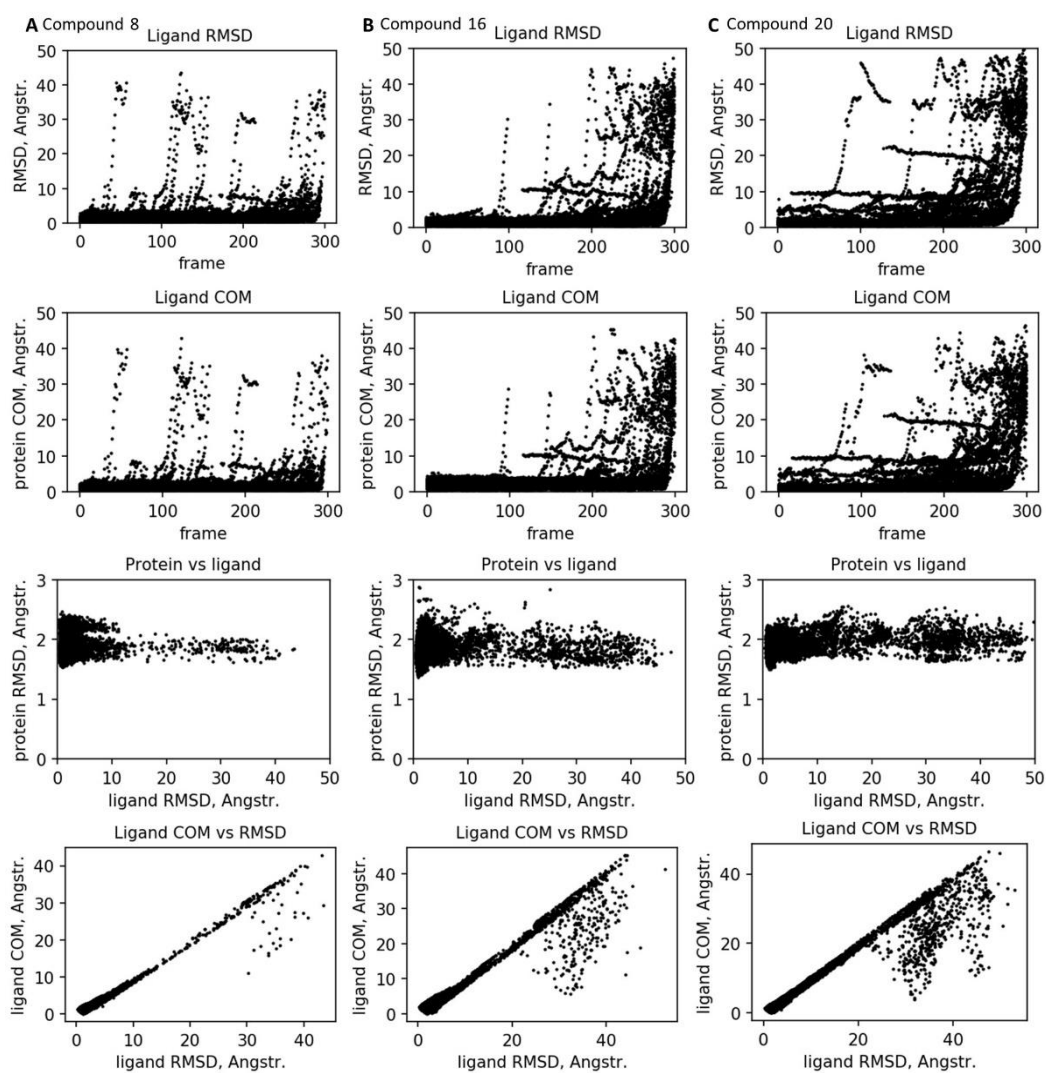

***Figure S5*** *Structural variations in the last 300 frames of RAMD dissociation trajectories for three studied compounds of HSP90 (compounds 8, 16, and 20, see Fig.2 of the main text): RMSD and COM of the ligand relative to the bound state as a function of the simulation frame (plots in the first two rows),*

RMSD of the protein versus RMSD of the ligand (third raow),  and ligand COM vs RMSD (lower plots). In all cases, the RMSD of heavy atoms was computed

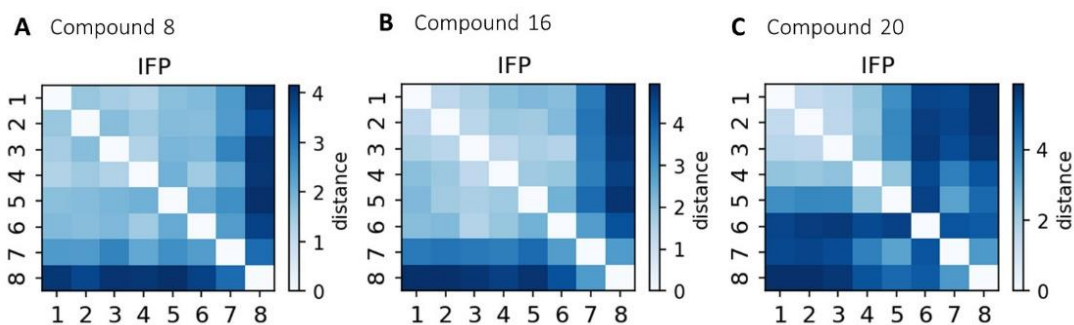

*Figure S6* 2D Euclidian distance matrix in the IFP space between cluster means for the clusters shown in *Figure 4* of the main text

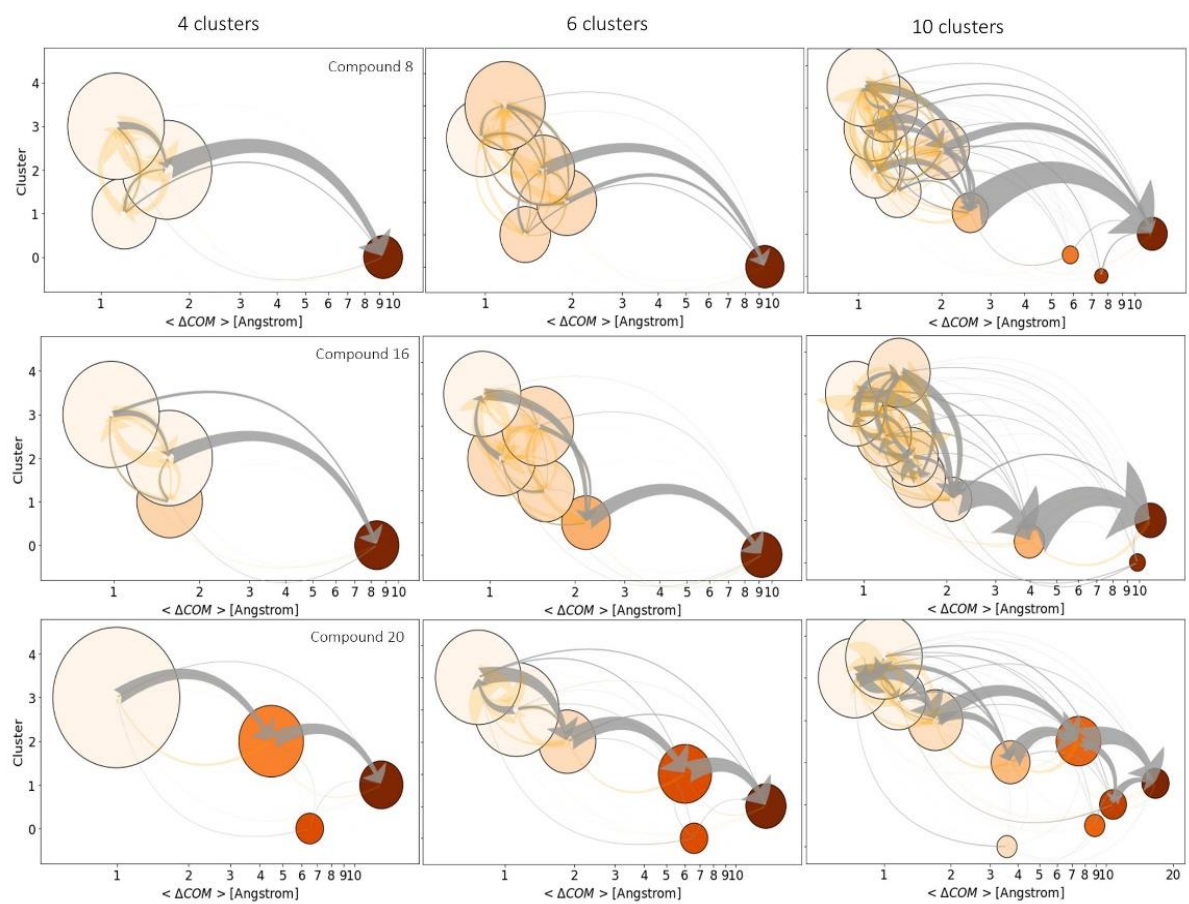

*Figure S7*. Illustration of the effect of cluster number on the structure of the simulated dissociation pathways. IFPs and nonspecific protein-ligand contacts were used for clustering. Plots are shown for specification of 4, 6 and 10 clusters in the k-means clustering of IFPs in RAMD trajectories for the egress of compounds 8, 16 and 20 from HSP90.

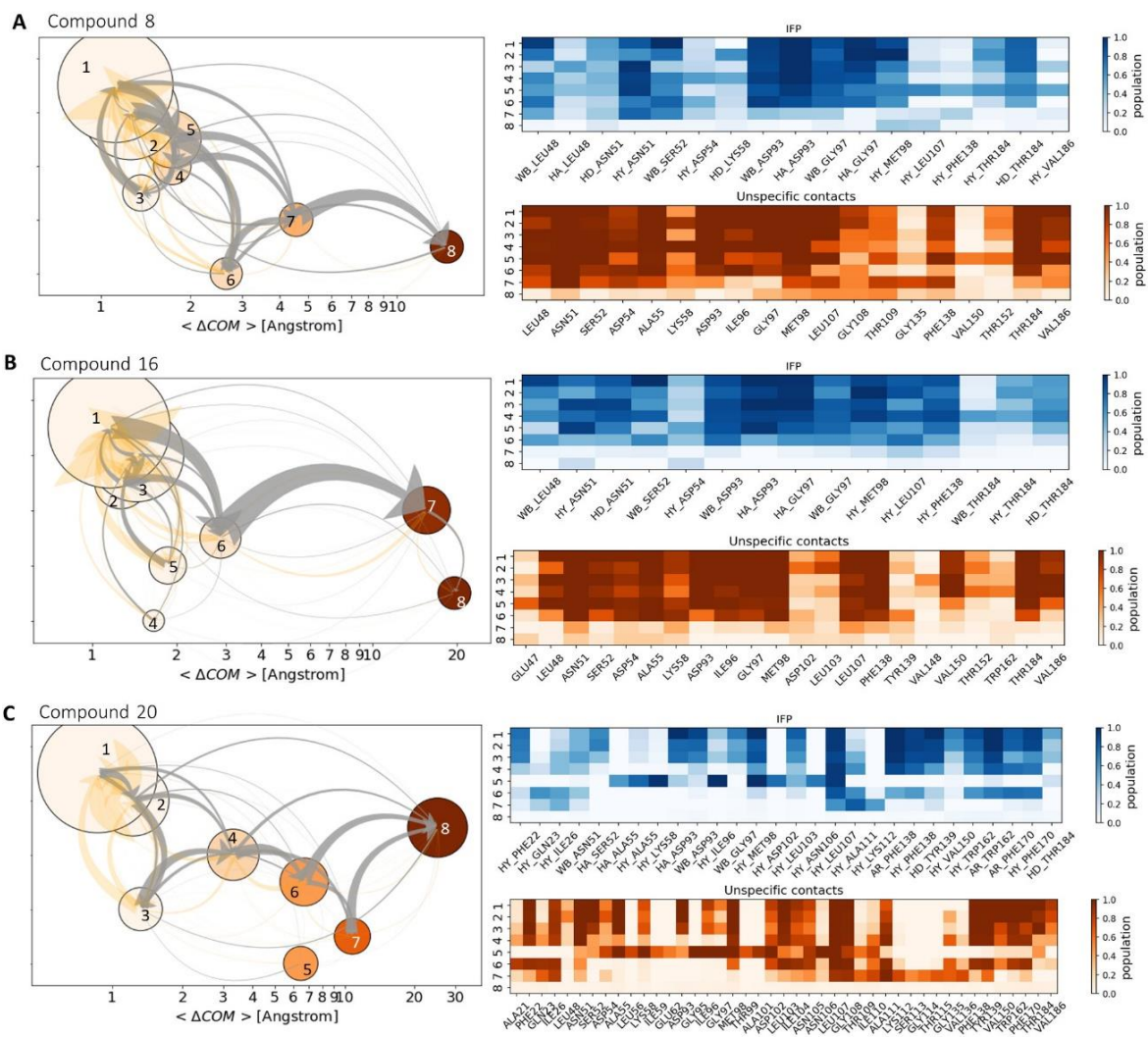

***Figure S8***. *Illustration of the clustering obtained by applying the Gaussian Mixture method to the RAMD dissociation trajectories of the 3 compounds considered in the manuscript (panels A, B, and C). IFPs and nonspecific protein-ligand contacts were used for clustering. The left panels show the observed dissociation pathways (see caption of **Figure 4** of the main text). The right panels illustrate the cluster composition in the space of IFPs (coloured in the blue pallet) and non-specific contacts (coloured in the orange pallet).*

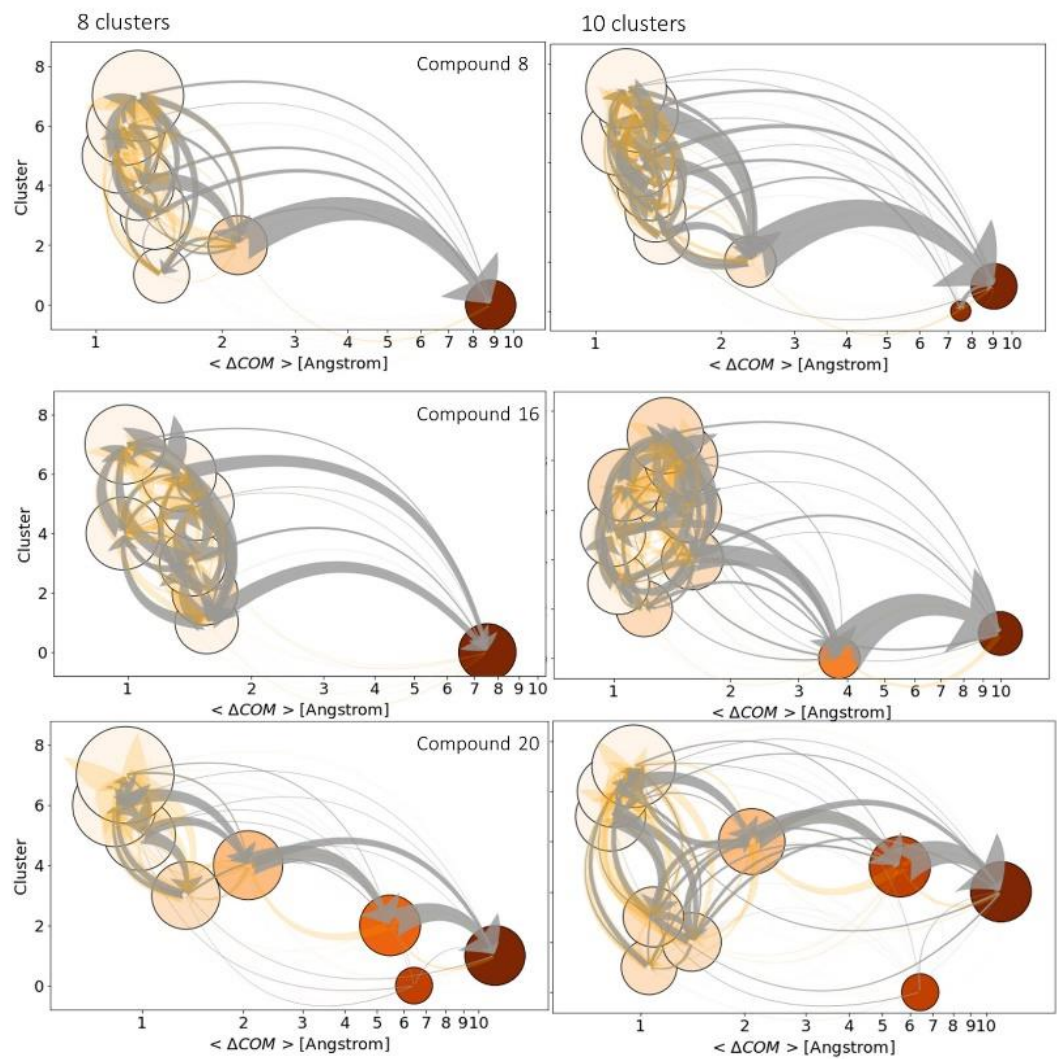

*Figure S9*. *Illustration of dissociation pathways of three compounds generated using the same procedure as described in the main text and in figure caption of **Figure 4** of the main text but using k-mean clustering based on the IFPs only; results of clustering with 8 and 10 clusters are shown.*